\documentclass[preprint,12pt]{aastex}

\def\msun{{\rm\,M_\odot}} 

\def\zsun{{\rm\,Z_\odot}}
\newcommand{\etal}{et al.\ }

\newcommand{\mpc}{\, {\rm Mpc}}

\newcommand{\lya}{Ly$\alpha$ }

\def\h2{${\rm\,H_2}$}

\slugcomment{Submitted to ApJ}

\begin{document}

\title{Galaxies Inside Str\"omgren Spheres of Luminous Quasars at $z>6$: Detection of The First Galaxies}

\author{Renyue Cen\altaffilmark{1}}

\altaffiltext{1} {Princeton University Observatory, 
Princeton University, Princeton, NJ 08544; cen@astro.princeton.edu}

\accepted{ }

\begin{abstract}

The intrinsic \lya emission lines of
normal galaxies before reionization are much absorbed 
by the damping wing of the Gunn-Peterson trough,
rendering their direct detection nearly impossible,
if their intrinsic line widths are less than $\sim 100$km/s.
High redshift luminous quasars 
prior to the completion of cosmological reionization at $z\sim 6$,
on the other hand, 
are capable of producing large 
HII regions around them (Str\"omgren spheres) 
to allow their intrinsic \lya emission lines 
to be transmitted without overwhelming absorption (Cen \& Haiman 2000).
We suggest that targeted observations 
at the Str\"omgren spheres of known luminous quasars at $z\ge 6$
would be able to detect \lya emission lines of 
galaxies inside the Str\"omgren spheres largely unattenuated.
A tunable, very narrowband filter of
${\Delta\lambda\over \lambda} \sim 0.1\%$ 
or a narrowband filter of ${\Delta\lambda\over \lambda} \sim 1\%$ 
with follow-up spectroscopic identifications will be required.
Such observations could directly observe
the sources of cosmological reionization including
possibly the Pop III galaxies at $z=6-20$ by JWST.
Possible applications include
determinations of the ionization state of the intergalactic
medium, the sizes of the Str\"omgren spheres, the ages
of the quasars, the luminosity function of high redshift galaxies
and its evolution,
the spatial distribution of galaxies and its evolution,
the biased distribution of galaxies around quasars
and the anisotropy of quasar emission.
Observations using Keck-class telescopes 
may already be made to enable a differentiation 
between a fully neutral and a 10\% neutral intergalactic medium
at $z>6$.

\end{abstract}


\section{Introduction}

Recent observations of high redshift 
quasars from the Sloan Digital Sky Survey (SDSS)
suggest that the cosmological reionization 
was completed at $z\sim 6$
(e.g., Fan \etal 2001; Becker \etal 2001, Barkana 2001; 
Djorgovski, Castro, \& Stern 2001;
Cen \& McDonald 2002; Litz \etal 2002),
implying a largely opaque universe beyond $z\sim 6$.
In the context of the standard cold dark matter model (Bahcall \etal 1999)
there is by now a clear consensus that 
galaxies of mass $10^8-10^9\msun$ are
largely responsible for the final cosmological reionization
(e.g., Barkana \& Loeb 2001; Madau 2002;
Wyithe \& Loeb 2003a,b;
Mackey, Bromm, \& Hernquist 2003;
Cen 2003a,b; Venkatesan, Tumlinson, \& Shull 2003;
Ciardi, Ferrara, \& White 2003;
Somerville \& Livio 2003;
Fukugita \& Kawasaki 2003;
Sokasian \etal 2003).
The recent Wilkinson Microwave Anisotropy Probe (WMAP)
observations (Kogut \etal 2003)
appear to confirm a much more extended
reionization process, with the first reionization occurring
at a redshift possibly as high as $z=15-25$ (Cen 2003a).
As a result,
our observational frontier is quickly 
being pushed towards high redshifts
not thought possible just a few years ago.
It is of great interest to directly detect the sources of reionization
and the ionization state of the intergalactic medium
from $z=6$ to $z=30$.

To date the observed Lyman alpha emitters (LAE) at $6<z<6.6$
appear to be quite luminous
with star formation rate (SFR)
probably in excess of $40\msun$yr$^{-1}$ 
(Hu \etal 2002; Kodaira \etal 2003),
which, if assuming a starburst of duration of $5\times 10^7$yrs and
star formation efficiency of $20\%$,
would give $\ge 1\times 10^{10}\msun$ of stars.
The masses of these galaxies most likely exceed $1\times 10^{11}\msun$.
These observed LAEs are thus much more massive
than the typical galaxies thought to produce
most of the ionizing photons at $z\ge 6$ --
current observations are just scraping the tip of an iceberg.
Because of the damping wing of the Gunn-Peterson trough
cast by the significantly neutral intergalactic medium (IGM)
before reionization (Miralda-Escud\'e 1998), 
the intrinsic luminosities (after correcting for possible
gravitational lensing effects)
for these observed LAEs may be still larger.
Consistent with this, typical LAEs observed at somewhat lower redshift
appear to be less luminous by a factor of ten or more 
(Yan, Windhorst, \& Cohen 2003; Rhoads \etal 2003;
Taniguchi \etal 2003).
For smaller galaxies of mass
$10^8-10^9\msun$ typically with
an intrinsic \lya line width narrower than $100$km/s,
the absorption by the damping wing
of intervening neutral IGM is dramatically 
more severe, probably rendering them undetectable (Haiman 2002).

In this {\it Letter} 
we suggest that targeted, very narrowband observations
of \lya emission at Str\"omgren spheres of known luminous 
quasars at $z\ge 6$ may provide a unique
window to directly detect the bulk
of the sources for cosmological reionization,
because a large fraction of galaxies in such regions can transmit
their \lya emission lines largely unattenuated.

%
%
%
%

\section{Galaxies Inside Str\"omgren Spheres of Luminous Quasars}

In the absence of hydrogen recombination with a static density field,
the proper radius $R_S$ of the Str\"omgren sphere of a
quasar of age $t_Q$ is such that the number of hydrogen atoms in the sphere
equals the total number of ionizing photons produced by the source
(Cen \& Haiman 2000):
\begin{eqnarray}
R_S = \left[ \frac{3\dot N_{\rm ph} t_Q}{4\pi x_{\rm IGM} \langle n_H \rangle}\right]^{1/3}=4.3x_{\rm IGM}^{-1/3}\left(\frac{\dot N_{\rm ph}}{1.3\times 10^{57}{\rm s^{-1}}}\right)^{1/3}\left(\frac{t_Q}{2\times10^7{\rm yr}}\right)^{1/3}\left(\frac{1+z_Q}{7.28}\right)^{-1}~{\mpc},
\end{eqnarray}
\noindent
where $\langle n_H \rangle$ is
the mean hydrogen density within $R_S$, 
$x_{\rm IGM}$ is the neutral hydrogen fraction of the IGM,
and $\dot N_{\rm ph}$ is emission rate
of ionizing photons from the quasar.  For $\langle n_H \rangle$, we have
adopted the mean IGM density with $\Omega_{\rm b}=0.047$, and $\dot
N_{\rm ph}=1.3\times 10^{57}~{\rm s^{-1}}$ 
is the observed flux of SDSS 1030+0524 (Haiman \& Cen 2002).
In the realistic case with an evolving cosmological density field 
and a finite hydrogen recombination time, 
$R_S$ can be computed in detail (Cen \& Haiman 2000).
But $R_S$ estimated in equation (1)
gives a reasonable approximation to the truth (Haiman \& Cen 2002)
at redshift $z\le 8$.

At sufficiently large separations 
from the \lya line center (${\Delta\lambda\over \lambda} \gg 2\times 10^{-8}$)
the optical depth of the damping wing is (Miralda-Escud\'e 1998)
\begin{equation}
\tau_d(\Delta\lambda)= 1.3\times 10^{-3} x_{\rm IGM} {\Omega_b h (1-Y)\over 0.03}{H_0(1+z)^{3/2}\over H(z)}({1+z\over 6})^{3/2}({\Delta\lambda\over \lambda})^{-1},
\end{equation}
\noindent
where $\Omega_b$ is baryon density parameter,
$Y$ is the helium mass fraction of the primordial gas,
$H_0$ and $H(z)$ are the Hubble constant at redshift zero and $z$,
respectively,
$\Delta\lambda$ and $\lambda$ are the observed wavelength separation
from the line center and the observed wavelength of the \lya line,
respectively.
Replacing ${\Delta\lambda/\lambda}$
by $H(z) \Delta r/c$, 
where $\Delta r$ is the distance between the surface of the 
Str\"omgren sphere and the point $\vec r$ inside the 
Str\"omgren sphere in question along the line of sight
and $c$ is the speed of light,
and approximating $H(z)\approx H_0(1+z)^{3/2}\Omega_M^{1/2}$,
we obtain 
\begin{equation}
\tau_d(\vec r)= 1.2 x_{\rm IGM} ({\Omega_M\over 0.27})^{-1} ({\Omega_b\over 0.047}) ({\sqrt{R_S^2-r^2\sin^2\theta}-r\cos\theta\over 1\mpc})^{-1},
\end{equation}
\noindent
where $r$ is the proper quasar-centric distance of the point
in question and $\theta$ is the angle between the line segment
from the quasar to the observer and the line segment from
the quasar to $\vec r$.

The Gunn-Peterson optical depth due to the residual
neutral hydrogen inside the Str\"omgren sphere,
in the absence of other ionizing sources,
can be expressed as (Cen \& Haiman 2000)
\break
$\tau_{r}(r)= 0.34({\Omega_M\over 0.27})^{-1/2} ({\Omega_b\over 0.047}) 
\left(\frac{\dot N_{\rm ph}}{1.3\times 10^{57}~{\rm s^{-1}}}\right)
({1+z\over 7.28})^{9/2}
({r\over 1\mpc})^{2}$,
due to a very small residual neutral fraction
$x_r\le 10^{-4}$.
As a result of small $x_r$,
the local mini-Str\"omgren spheres 
produced by \lya galaxies would render the absorption
due to all residual neutral hydrogen inside
the Str\"omgren sphere negligible.
The total \lya optical depth 
at $\vec r$ is hence just $\tau_d$. 
From Equation (3)
we see that, for a luminous quasar 
like SDSS 1030+0524,
\lya lines from galaxies located within
the central region region of $r\le 1$Mpc,
would suffer a total optical depth of $\tau_t=0.22-0.36$
corresponding $\theta$ from $180^o$ to $0^o$,
i.e., a flux reduction of $\le 31\%$,
assuming $x_{\rm IGM}=1.0$.
It is evident that there is a significant region
within the Str\"omgren sphere where
most of the \lya emission galaxies 
are able to transmit the \lya emission lines
largely unattenuated.

There are several potentially very interesting applications
from direct observations of regions around luminous quasars
prior to the completion of cosmological reionization.
Before turning to the possible applications it is useful
to estimate the abundance of possibly detectable galaxies
at $z\ge 6$.
Assuming that the fraction of all mass having formed into halos
with masses $M_h>10^9\msun$ 
is $f_g$
and that starbursts each with a duration of $t_{\rm sb}$ 
has occurred once in each halo in the last half Hubble time,
then we find that the comoving number density
of such galaxies is 
$n_g=1.2({t_{\rm sb}\over 5\times 10^7{\rm yrs}})
({f_g\over 0.10})({M\over 10^9\msun})^{-1}h^{3}$Mpc$^{-3}$.
Assuming a star formation efficiency $c_*$
we obtain the star formation rate in each galaxy
${\rm SFR}=0.68 ({c_*\over 0.20})({t_{\rm sb}\over 5\times 10^7{\rm yrs}})^{-1}({M\over 10^9\msun})\msun$/yr.
This simple estimate, however, involves two currently unknown 
parameters, $c_*$ and $t_{\rm sb}$.

\begin{figure}
\plotone{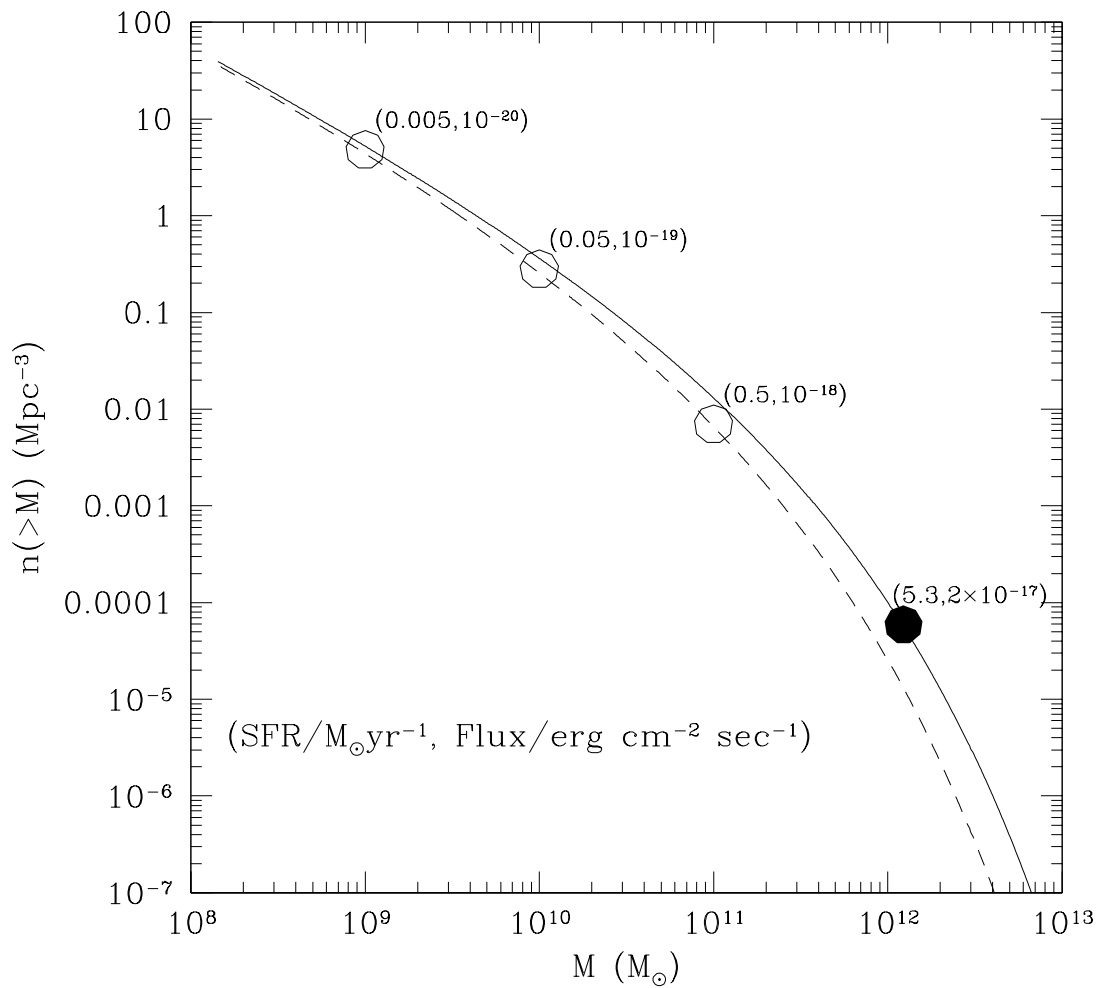}
\caption{
shows the cumulative number density of galaxies 
as a function of mass at $z=5.7$ (solid)
and $z=6.5$ (dashed)
in a cosmological constant dominated cold dark
matter model, consistent with WMAP observations,
with the following parameters:
$\Omega_M=0.29$, $\Lambda=0.71$, $H=70$km/s/Mpc,
$\sigma_8=0.80$ and $n=1.0$.
The solid dot is an observational data point
from Rhoads \etal (2003).
The open circles are three representative
galaxy masses with approximate star formation rates and 
\lya line fluxes indicated. 
}
\label{rat}
\end{figure}
 
Perhaps a better way to estimate the abundance and
luminosity of LAE is to utilize both
hierarchical CDM model and LAE survey observations at
highest accessible redshift, currently at $z\sim 5.7$.
Figure 1 shows the predicted cumulative
number density of galaxies at $z=5.7$ (solid curve)
and $z=6.5$ (dashed curve) for 
a cosmological constant dominated cold dark
matter model with 
$\Omega_M=0.29$, $\Lambda=0.71$, $H=70$km/s/Mpc,
$\sigma_8=0.80$ and $n=1.0$.
The observational data point for LAE
shown as the solid dot in Figure 1
is taken from Rhoads \etal (2003), who give
$n({\rm SFR}>5.3 \msun {\rm yr}^{-1})=5.9\times 10^{-5}$Mpc$^{-3}$
at $z\sim 5.7$.
We have ``normalized" the observational data point
to the theoretical prediction at $z=5.7$ by requiring 
the same abundance, which then gives an 
implied mass threshold of the observed LAE by Rhoads \etal
Clearly, even the LAE observed at $z\sim 5.7$ by Rhoads \etal 
are quite massive with masses about $10^{12}\msun$,
if the adopted cold dark matter model is correct.
The computed lower bound of star formation rate
for the observed LAE at $z\sim 5.7$
by Rhoads \etal of ${\rm SFR}>5.3 \msun {\rm yr}^{-1}$ corresponds to
the flux sensitivity of $2\times 10^{-17}$erg/cm$^2$/sec
of their LALA survey.
If we assume that the 
\lya luminosity of an LAE is simply proportional to its mass,
it would require a flux sensitivity of
$(10^{-18}, 10^{-19}, 10^{-20}$)erg/cm$^2$/sec
to detect galaxies of mass
of $(10^{11}, 10^{10}, 10^{9})\msun$
with corresponding abundance of
$(0.01, 0.3, 5.0)$Mpc$^{-3}$ at $z=6.5$ 
(open circles in Figure 1), respectively.

Between night sky lines there are a few good observational
windows at $\lambda > 0.79\mu m$ (corresponding to $z>5.5$
for the \lya line), most notably the one center
at $z\sim 5.7$ and the other at $z\sim 6.5$.
However, even within those ``good" windows the required
integration time to detect fainter LAE
using Keck telescopes is quite long,
$t=({\rm flux}/10^{-17}{\rm erg/cm}^2$/{\rm sec})$^{-2}~$hours
(Gjorgovski 2003, private communications;
Fan 2003, private communications),
implying that approximately a hundred hours
is needed on Keck to detect 
galaxies of mass $10^{11}\msun$,
{\it even in the absence of significantly attenuation by neutral IGM}.
The current observational limit for \lya flux
roughly corresponds to ${\rm SFR}\sim$ a few $\msun/$yrs at $z\sim 6$ 
(Hu \etal 2002;
Yan, Windhorst, \& Cohen 2003; Rhoads \etal 2003;
Taniguchi \etal 2003).
Detecting even lower mass LAE may be quite 
infeasible using Keck telescopes.
Furthermore, for ground-based 
observations where the primary background noise in the interested
spectral regime comes from sky background,
it is a very tedious task to search for 
usable windows between numerous strong sky lines to 
find high redshift galaxies and quasars.
Hence, it is likely necessary to resort to James Webb Space Telescope (JWST)
for detecting
the bulk of galaxies (of mass $M_h\sim 10^9\msun$) at $z>6$.
As pointed out by Oh (1999), JWST should be able to detect
the H$_\alpha$ line (and possibly H$_\beta$ and other Balmer lines)
of galaxies of mass $M_h\sim 10^9\msun$ at $z>6$.
A joint detection in 
\lya and H$_\alpha$ would not only firmly
determine the redshift of each galaxy but also
shed very useful light on important physical
processes of each galaxy including dust content.
In addition, the ratio of 
\lya to H$_\alpha$ lines would provide a cross-check
of the damping wing modulated spatial distribution 
of LAE or a tool to break possible degeneracies 
in the spatial distribution of LAE from \lya line alone.

We note that three factors could boost the effectiveness of star formation 
at high redshift in terms of \lya emission.
First, the gas at high redshift may be much metal poor;
for example, taking the same Salpeter IMF would give about
a factor of $2$ more ionizing photons hence \lya photons
for stars with $Z\approx 10^{-2}\zsun$ compared
to stars with solar metallicity.
Second, if the metal poor gas at high redshift forms
relatively more massive stars, the efficiency
of producing ionizing photons would be higher.
Third, the dust content in the high-z metal-poor
galaxies should be lower,
possibly allowing for relatively more \lya photons to escape.
These factors may render
high redshift LAE more luminous that their low redshift
counterparts at a star formation rate.


\section{Applications}

\subsection{Probing the Ionization State of the IGM and the Size
of the Str\"omgren Sphere}

\begin{figure}
\epsscale{0.8}
\plotone{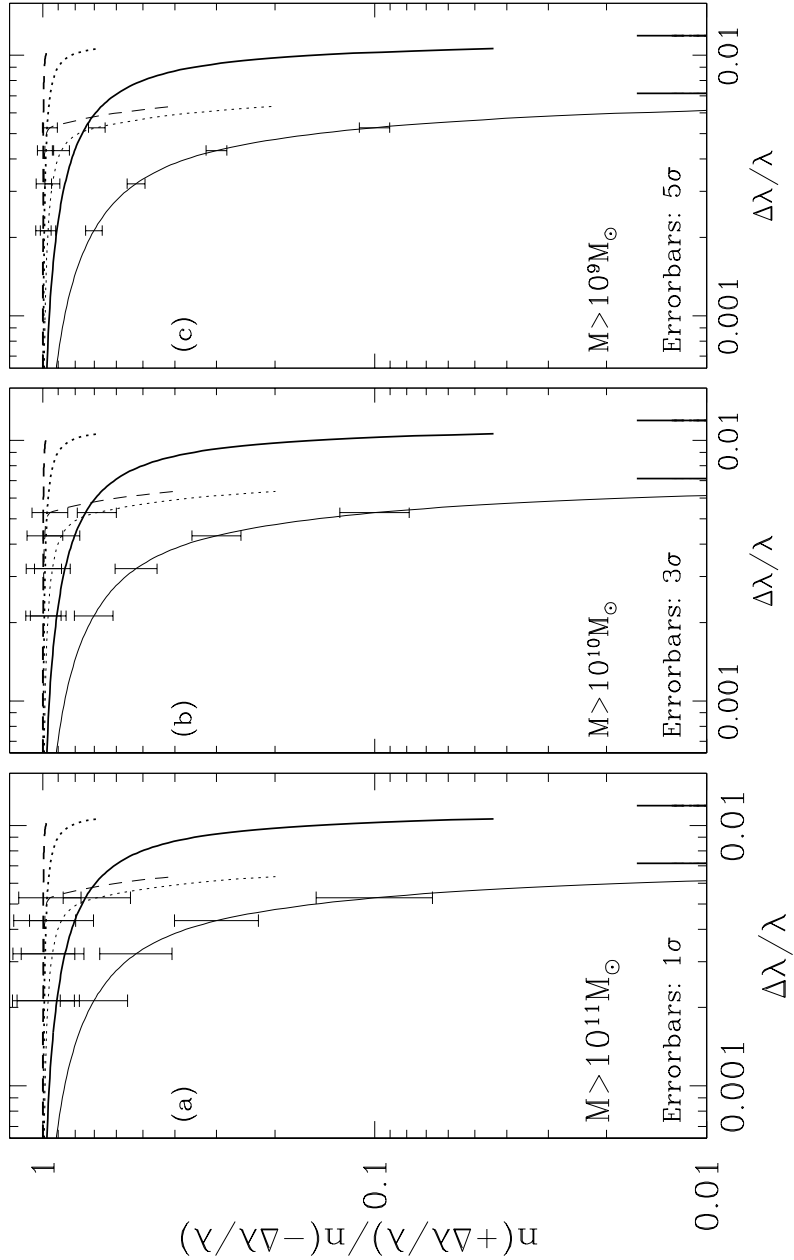}
\caption{
shows the ratio of the number of galaxies 
at a distance $\Delta\lambda/\lambda$ 
(with width $\delta{\Delta\lambda\over\lambda}$) 
from the quasar
on the blue side over
the number of galaxies 
at the same distance from the quasar on the red side within
a transverse radius of $r_p=2$Mpc proper,
as a function of $\Delta\lambda/\lambda$.
Three panels correspond to three flux sensitivities
of $10^{-18}$erg/cm$^{2}$/sec (Panel a), $10^{-19}$erg/cm$^{2}$/sec (Panel b)
and $10^{-20}$erg/cm$^{2}$/sec (Panel c), 
which are then translated into the threshold
mass of detectable galaxies using Figure 1.
In each panel two sets of cases are shown.
Case (1): $R_s=5$Mpc proper with $x_{\rm IGM}=(1.0,0.1,0.01)$,
respectively, shown as solid (green), dotted (red) and dashed (blue).
Case (2): $R_s=3$Mpc proper with $x_{\rm IGM}=(1.0,0.1,0.01)$.
Each set asymptotically curves downward towards
the radius of the respective Str\"omgren sphere indicated
by the vertical tick sitting on the bottom x-axis.
The errorbars are statistical.
}
\label{rat}
\end{figure}

Figure (2) shows the ratio of the number of galaxies 
at a distance $\Delta\lambda/\lambda$ 
(with width $\delta{\Delta\lambda\over\lambda}$) 
from the quasar on the blue side over
the number of galaxies 
at the same distance from the quasar on the red side within
a quasar-centric transverse radius of $r_p=2$Mpc proper,
as a function of $\Delta\lambda/\lambda$.
Three panels correspond to three flux sensitivities
of $10^{-18}$erg/cm$^{2}$/sec (Panel a), $10^{-19}$erg/cm$^{2}$/sec
(Panel b) and $10^{-20}$erg/cm$^{2}$/sec (Panel c), 
which are then translated into the threshold
masses of detectable galaxies using Figure 1 to infer the 
corresponding abundances.
In each panel two sets of cases are shown.
Case (1): $R_s=5$Mpc proper with $x_{\rm IGM}=(1.0,0.1,0.01)$,
respectively, shown as thick solid, thick dotted and thick dashed curves.
Case (2): $R_s=3$Mpc proper with $x_{\rm IGM}=(1.0,0.1,0.01)$,
respectively, shown as thin solid, thin dotted and thin dashed curves.
We choose to plot the ratio instead of the direct number density
to avoid ambiguity  
due to a possibly non-uniform, 
quasi spherical distribution of galaxies around quasars.
Galaxies will still likely be clumpy and clustered on small scales
but a large volume considered 
would smooth out most of the fluctuations.

Let us now estimate the accuracy of the data points.
The comoving volume enclosed by a cylindrical region
within a transverse radius $r_p=2\mpc$ 
and a line-of-sight distance of 
$\Delta\lambda/\lambda=0.01$ 
(from $\Delta\lambda/\lambda=-0.005$ 
to $\Delta\lambda/\lambda=0.005$)
is $2.1\times 10^4\mpc^3$ comoving at $z=6.28$,
which is then multiplied by the number density of galaxies
using the respective $n$ from Figure 1 to yield 
the corresponding total number of galaxies above 
the respective mass threshold.
Five data points are used for the indicated $\Delta\lambda/\lambda$ range.
The errorbars are shown in each panel 
for the case with $R_s=3$Mpc proper.
From Panel (a) of Figure 2 
it is clear that $x_{\rm IGM}=1$ and $x_{\rm IGM}=0.1$ 
are distinguishable at $2-3\sigma$ level
even for galaxies with mass $M>10^{11}\msun$ (Panel a),
detectable by Keck telescope with an integration time of order of
$100$ hours for each galaxy (assuming one spectrum 
per observation) with total $\sim 210$ galaxies.
This might be a costly project 
but the potential payoff is attractive.
Cases with $x_{\rm IGM}=0.1$ and $x_{\rm IGM}=0.01$ 
can only be marginally differentiated at this sensitivity level.
At the same time, it is evident that
the models with different $R_s$ can be more easily
distinguished, allowing for a precise measurement of $R_s$.
In addition, a precise measurement
of $R_s$, in conjunction with
the spatial information (see \S 3.2),
may allow a determination of the age of the quasar.
With JWST, if one can reach flux levels 
$10^{-20}-10^{-19}$erg/cm$^2$/sec,
very precise measurements of $x_{\rm IGM}$ can be made,
as can be seen from Panels (b,c) in Figure 2.

We note that $\Delta\lambda/\lambda=0.005$ corresponds to 
a line-of-sight comoving distance from the quasar of $16$Mpc at $z=6.28$.
The gravitational influence of the quasar host galaxy
at that distance is likely to be negligible,
alleviating some possible complications including infall.
In addition, the typical peculiar velocities of galaxies at high redshift
is likely be $<200$km/s, which
would correspond to a possible displacement of 
$\Delta\lambda/\lambda=6\times 10^{-4}$, too small to significantly
matter.

Finally, we note that, in the case of a lone LAE (not sitting
inside a large Str\"omgren sphere of a luminous quasar),
an LAE estimated to have  
an SFR about an order of magnitude higher
than the solid dot in Figure 1 (Hu \etal 2002)
can only transmit a small fraction of \lya flux
due to attenuation by the neutral IGM (Haiman 2002).
Lower mass LAE are further devastated by their lower
intrinsic line widths in addition to smaller Str\"omgren spheres.
As Haiman (2002) pointed out,
no \lya line of an LAE narrower than
$100$km/s can be observed.
Therefore, it is estimated that no LAE galaxies
less massive than $10^{10}-10^{11}\msun$ will 
be detectable in a random field outside 
a large Str\"omgren sphere of a luminous source.

\subsection{Galaxy Luminosity Function and Spatial Distribution}

Fitting the data to obtain $R_s$ and $x_{\rm IGM}$ (\S 3.1)
would yield precise information on $\tau_d$, which
in turn would allow a determination of a region
on the red side of the quasar where attenuation
due to intervening neutral IGM 
is negligible.
Consequently, one can construct an unbiased luminosity function 
given the likely availability of up to thousands of galaxies
in that region, which would further 
allow for corrections to be made 
for ``missing" galaxies in other regions where $\tau_d$ is appreciable.
Moreover, the spatial distribution of galaxies
in this region and others in other regions
would allow an accurate determination of the 
clustering properties of galaxies 
up to a scale of $10-30$Mpc comoving.

Given the determined luminosity function and $x_{\rm IGM}$,
along with follow-up spectroscopic observations of 
the galaxies detected which would produce the line widths for these
galaxies,
we could self-consistently determine 
a fraction of these galaxies that would be directly
observed even sitting in an average region
and their possible observable features.

\subsection{Biased Distribution Around Quasars}

Taking the farthest red region as an unbiased region,
this would then allow a determination of 
luminosity-dependent biased distributions of galaxies,
down to the lowest luminosity 
of galaxies in the farthest red region.
This may shed useful light on the formation 
of quasars and their host galaxies.
Whether or not the farthest red region 
is an unbiased, ``average" region can be checked
using the same data, where one may expect to see
a diminished radial gradient.

\subsection{Probing Anisotropy of Quasar Emission}

The three-dimensional information of the distribution 
of galaxies around a quasar makes it possible
to probe the angular distribution of emission
from the quasar. 
It should be relatively straightforward
to build self-consistent models to account for
possible outcomes from observations.
For example, if galaxy abundance drops 
off rapidly at a transverse distance significantly
smaller than the inferred $R_s$,
it would indicate that the quasar emission
is bi-polar, whose angles can be inferred.
If, in addition, there is a strong suppression
of galaxies on the red side of the quasar even
at small $r_p$,
it would imply that the quasar emission is largely beamed
towards us.
We will not attempt to make more detailed models here.
If the standard unified model for AGN/quasars (e.g., Antonucci 1993)
is correct,
the ensuing consequence on the distribution
of LAEs around quasars will clearly be detectable.
This method may provide the first and accurate way to test this picture.

In summary, we show that, with a tunable,
very narrowband filter of 
$\Delta\lambda/\lambda\sim 0.1\%$ 
or a narrowband filter of ${\Delta\lambda\over \lambda} \sim 1\%$ to
survey a luminous quasar Str\"omgren sphere
with follow-up spectroscopic identifications
of the \lya emission lines of individual galaxies
inside the quasar Str\"omgren sphere,
we could obtain a wealth of information for the
high redshift galaxies that are responsible
for reionizing the universe at $z\sim 6$.
This same technique may be applicable to quasars
at any high redshift.
At higher redshift, galaxies are somewhat less massive,
but they may have a more top-heavy IMF,
possibly a higher star formation efficiency,
a shorter starburst duration and much lower dust content.
These positive, compensatory effects could possibly render 
high redshift galaxies with 
mass $10^7-10^8\msun$ intrinsically as luminous as
galaxies of mass $10^8-10^9\msun$ at $z=6$.
We expect that JWST should be able to detect
these galaxies at $z>20$ (Eisenstein 2003),
provided that a tunable, narrowband filter at appropriate
wavelengths is available.

\section{Conclusions}

We show that targeted observations at 
the Str\"omgren spheres of known luminous quasars
with tunable very narrowband filters of width $\sim 0.1\%$ 
or a narrowband filter of ${\Delta\lambda\over \lambda} \sim 1\%$ to
survey a luminous quasar Str\"omgren sphere
with follow-up spectroscopic identifications
of the \lya emission lines of individual galaxies
would provide a direct way to observe the \lya emission lines
of the bulk of galaxies at $z>6$
in the mass range $10^8-10^9\msun$ that are believed
to be responsible for the second cosmological reionization.
At present it appears that ground-based observations, such as
by Keck, can only detect a relatively small fraction, high end
tail of galaxies with mass $>10^{11}\msun$. 
But even such observations 
may already be able to differentiate 
between a fully neutral and a 10\% neutral intergalactic medium 
at $z>6$.
JWST should provide the unique opportunity
to detect the bulk of galaxies at high redshift.
The likely gain in sensitivity with a tunable, very narrowband filter 
may allow JWST to detect the Pop III galaxies 
of mass $10^7-10^8\msun$ with a top-heavy IMF at $z>20$.
In this case, the key is to search for luminous enough quasars 
at these very high redshifts, perhaps requiring an infrared survey 
comparable to SDSS. This is presently technologically challenging
but the potential scientific payoff is immense.

We show that such observations will allow for accurate
determinations of the ionization state of the IGM,
the sizes of the Str\"omgren spheres, the ages
of the quasars, the luminosity function of high redshift galaxies
and its evolution,
the spatial distribution of galaxies and its evolution,
the biased distribution of galaxies around quasars
and the anisotropy of quasar emission.

If the universe was indeed reionized twice (Cen 2003a),
then there will be a window of opportunity somewhere
at $z=10-25$, where galaxies can transmit \lya lines
(as well as optical-UV continuum and other emission lines)
unabsorbed; they should be directly observable by JWST 
even in the absence of luminous quasars.
A signature continuum cutoff of the spectra of these galaxies 
shortward of $1216(1+z_{rec,2})\AA$ would pin down
the redshift of the second cosmological recombination epoch $z_{rec,2}$
and then a rapid thinning out of the number density 
of \lya galaxies at some redshift
$z>z_{rec,2}$ would be indicative of the first reionization epoch.

\acknowledgments
This research is supported in part by NSF grant AST-0206299.
I thank George Djorgovski, Xiaohui Fan and Rogier Windhorst for useful comments
and discussions, Michael Strauss for pointers to
useful astronomy websites,
and the referee for useful suggestions which significantly
strengthen the paper.

\end{document}